\documentclass[conference]{IEEEtran}
\usepackage{amsthm}
\usepackage{amsmath}
\usepackage{amssymb}
\usepackage{subfigure}
\usepackage{epic}

\newcommand{\Nb}{{\mathbb{N}}}

\newcommand{\B}{{\cal B}}

\newcommand{\R}{{\cal R}}

\newcommand{\N}{{\cal N}}
\newtheorem{prop}{Proposition}

\theoremstyle{remark}

\newtheorem{example}{Example}

\begin{document}

\title{Weak randomness and Kamae's theorem on normal numbers}
\author{\IEEEauthorblockN{Hayato Takahashi}
\IEEEauthorblockA{The Institute of Statistical Mathematics,\\
10-3 Midori-cho, Tachikawa, Tokyo 190-8562, Japan.\\
 e-mail: hayato.takahashi@ieee.org.}}
 \markboth{}{}
 \maketitle
\begin{abstract}
A function from sequences to their subsequences is called selection function. 
A selection function is called admissible (with respect to normal numbers) if for all normal numbers, their subsequences obtained by the selection function are normal numbers.
 In Kamae (1973)  selection functions that are not depend on sequences (depend only on coordinates) are studied, and their necessary and sufficient condition for admissibility is given.
In this paper we introduce a notion of weak randomness and study an algorithmic analogy to the Kamae's theorem. 
\end{abstract}
\IEEEpeerreviewmaketitle

\section{Introduction}\label{sec-intro}
In this paper we study subsequences of random numbers.
A function from sequences to their subsequences is called selection function. 
A selection function is called {\it admissible} (with respect to normal numbers) if for all normal numbers, their subsequences obtained by the selection function are normal numbers.
 In Kamae \cite{kamae73}  selection functions that are not depend on sequences (depend only on coordinates) are studied, and their necessary and sufficient condition for admissibility is given.
In this paper we introduce a notion of weak randomness and study an algorithmic analogy to the Kamae's theorem.

Let \(\Omega\) be the set of infinite binary sequences.
For \(x,y\in\Omega\), let \(x=x_1x_2\cdots, y=y_1y_2\cdots,\ \forall i\ x_i, y_i\in\{0,1\}\).
Let \(\tau:\Nb\to\Nb\) be a strictly increasing function such that 
\(\forall i\ (y_i=1\leftrightarrow\exists j~i=\tau(j))\).
If \(\sum_i y_i=n\) then \(\tau(j)\) is defined for \(1\leq j\leq n\).
For \(x,y\in\Omega\) let \(x/y\) be the subsequence of \(x\) selected at \(y_i=1\), i.e.,
\(x/y=x_{\tau(1)}x_{\tau(2)}\cdots\).
For example, if \(x=0011\cdots,\ y=0101\cdots\) then \(\tau(1)=2, \tau(2)=4\) and \(x/y=01\cdots\).
Let \(x_1^n:=x_1\cdots x_n\) and \(y_1^n:=y_1\cdots y_n\).
Then \(x_1^n/y_1^n\) is defined similarly. 

In Kamae \cite{kamae73}, it is shown that the following two statements are equivalent under the assumption that \(\liminf \frac{1}{n}\sum_{i=1}^n y_i>0\):\\
(i) \(h(y)=0\).\\
(ii) \(\forall x\in \N\ x/y\in \N\),\\
where \(h(y)\) is  Kamae entropy \cite{{brudno83},{lambalgen87}} and \(\N\) is the set of binary normal numbers.
Roughly speaking, \(h(y)\) is the least upper bound of the entropy of the limit points (in the weak topology) of \(\frac{1}{n}\sum_1^n \delta_{T^i y}\), where \(\delta_x\) is 1 at \(x\) and 0 else, and \(T\) is shift.
If \(h(y)=0\), it is called completely deterministic, see \cite{{kamae73},{weiss71},{weiss00}}.
The part (i)\(\Rightarrow\) (ii) is appeared in \cite{weiss71}.

As a natural analogy,  the following equivalence (algorithmic randomness version of Kamae's theorem) under a suitable restriction on \(y\) is questioned  in Lambalgen \cite{lambalgen87},\\
(i) \(\lim_{n\to\infty}K(y_1^n)/n=0.\)\\
(ii) \(\forall x\in\R\ x/y\in\R\),\\
where \(K\) is the prefix Kolmogorov complexity and \(\R\) is the set of Martin-L\"of random sequences with respect to the uniform measure (fair coin flipping), see \cite{LV2008}.

In this paper, we show a similar equivalence for weak randomness. 
Let \(S\) be the set of finite binary strings. 
For \(x\in S\) let \(\Delta(x):=\{x\omega | \omega\in\Omega\}\), where \(x\omega\) is the concatenation of \(x\) and \(\omega\).
Let \((\Omega,\B, P)\) be a probability space, where \(\B\) is the sigma-algebra generated by \(\Delta(x), x\in S\).
We write \(P(x):=P(\Delta(x))\).
\(P\) is called computable if there is a computable function \(A\) such that \(\forall x, k\ |P(x)-A(x,k)|<1/k\).
We say that \(y\) is weakly random with respect to a computable \(P\) if 
\begin{equation}\label{weak}
\lim_{n\to\infty} K(y_1^n)/n=\lim_{n\to\infty}-\frac{1}{n}\log P(y_1^n),
\end{equation}
where the base of logarithm is 2.
For example if \(P\) is the uniform measure, i.e., \(P(x)=2^{-|x|}\) for all \(x\), where \(|x|\) is the length of \(x\), then
\(y\) is weakly random with respect to \(P\) if \(\lim_{n\to\infty} K(y_1^n)/n=1\).
If \(y\) is Martin-L\"of random sequences with respect to a computable ergodic \(P\)  then
from upcrossing inequality for the Shannon-McMillan-Breiman theorem \cite{hochman2009}, the right-hand-side of (\ref{weak}) exists (see also \cite{vyugin98}) and from
 Levin-Schnorr theorem \cite{LV2008} we see that (\ref{weak}) holds i.e., \(y\) is weakly random. 

\begin{prop}\label{main}
Suppose that \(y\) is weakly random with respect to a computable measure and \(\lim_n \frac{1}{n}\sum_{i=1}^n y_i>0\).
Then the following two statements are equivalent:\\
(i) \(\lim_{n\to\infty}K(y_1^n)/n=0.\)\\
(ii) \(\forall x\ \lim_{x\to\infty}K(x_1^n)/n=1,\)\\  \(\lim_{n\to\infty} \frac{1}{|x_1^n/y_1^n|}K(x_1^n/y_1^n |y_1^n )=1.\)
\end{prop}
Sketch of proof)\\
(i) \(\Rightarrow\) (ii)\\
Let \(\bar{y}:=\bar{y}_1\bar{y}_2\cdots\in\Omega\) such that \(\bar{y}_i=1\) if \(y_i=0\) and \(\bar{y}_i=0\) else for all \(i\).
Since 
\[ | K(x_1^n)-K(x_1^n | y_1^n) | \leq K(y_1^n)+O(1)\]
 and 
 \begin{align*}
& K(x_1^n | y_1^n)=K(x_1^n/ y_1^n, x_1^n/ \bar{y}_1^n | y_1^n) +O(1),
\end{align*}
if \(\lim_{n\to\infty}K(y_1^n)/n=0\) then 
we have
\begin{align*}
& \lim_{x\to\infty}K(x_1^n)/n=1\\
& \Rightarrow \lim_{n\to\infty} \frac{1}{n} K(x_1^n/ y_1^n, x_1^n/ \bar{y}_1^n | y_1^n)=1\\
& \Rightarrow \lim_{n\to\infty} \frac{1}{n}(K(x_1^n/ y_1^n| y_1^n)+K(x_1^n/ \bar{y}_1^n | y_1^n))=1\\
& \Rightarrow \lim_{n\to\infty} \frac{1}{|x_1^n/y_1^n|}K(x_1^n/y_1^n |y_1^n )=1.
\end{align*}
(ii) \(\Rightarrow\) (i)\\
Suppose that 
\[\lim_{n\to\infty} K(y_1^n)/n=\lim_{n\to\infty}-\frac{1}{n}\log P(y_1^n)>0,\]
for a computable \(P\).
Let \(f(n)\) be the least integer greater than \(-\log P(y_1^n)\).
Then by considering arithmetic coding, there is \(z=z_1z_2\cdots\in\Omega\) such that 
\(K(y_1^n)=K(z_1^{f(n)})+o(n)=f(n)+o(n)\) and \(z_1^{f(n)}\) is computable from \(y_1^n\).
Then we have \(\lim_{n\to\infty} K(z_1^n)/n=1\) and \(\limsup_{n\to\infty} \frac{1}{|z_1^n/y_1^n|}K(z_1^n/y_1^n | y_1^n)<1\).
\qed

\begin{example}
Let \(y\) be a computable sequence and \(P\) be a probability that has probability one at \(y\).
Then \(P\) is computable and (\ref{weak}) holds.
Therefore if \(\lim_n \frac{1}{n}\sum_{i=1}^n y_i>0\) and \(y\) is computable, it satisfies the condition of Proposition~\ref{main}. 
In particular Champernowne sequence satisfies the condition of the proposition and (i) holds, however its Kamae-entropy is not zero. 
\end{example}

\begin{example}
If \(y\) is a Sturmian sequence generated by an irrational rotation model with a computable parameter then \(y\) satisfies the condition of the proposition and (i) holds. 
\end{example}

\begin{center}
{\bf Acknowledgement}
\end{center}
The author thanks Prof.~Teturo Kamae (Matsuyama Univ.) for discussions and comments.

\end{document}